\documentclass[preprint]{aastex}

\begin{document}
\title{Strong-Field Electrodynamics} 
\author{Andrei Gruzinov}
\affil{CCPP, Physics Department, New York University, 4 Washington Place, New York, NY 10003}

\begin{abstract}

Strong-Field Electrodynamics (SFE) is Maxwell theory with a  certain Lorentz-covariant Ohm's law which uses only the electromagnetic degrees of freedom. We show that SFE is {\it semi-dissipative}: while the dissipation rate of the electromagnetic energy is non-negative, it can be exactly zero for non-trivial electromagnetic fields. 

It appears that SFE is well-defined for arbitrary electromagnetic fields. It should be possible to calculate the dissipative pulsar magnetosphere and resolve the magnetic separatrix using SFE.

We show that SFE reduces to Force-Free Electrodynamics (FFE) in the large conductivity limit. In the regions where the ideal FFE 4-current is space-like, SFE predicts small dissipative corrections. In the regions where the ideal FFE 4-current is time-like, SFE predicts a zero correction. This indicates that bright pulsars radiate primarily from the magnetic separatrix.

\end{abstract}

\section{Introduction}

It has been argued that large-scale electromagnetic fields of bright Crab-like pulsars are properly described by Force-Free Electrodynamics (FFE) (Gruzinov 2007 a). This allows to calculate the ideal pulsar magnetosphere (Contopoulos, Kazanas, Fendt 1999, Gruzinov 2005, 2006, Spitkovsky 2006). 

FFE is dissipationless. In FFE the entire spin-down power is carried by the large-scale Poynting flux. Formally, FFE predicts zero radiation. This appears to be a good approximation for Crab-like pulsars -- observed luminosity is indeed much smaller than the estimated spin-down power. Radiation should come as a small dissipative correction to FFE.

But of course, we must find a way to calculate these small dissipative corrections, because most of the pulsar data concerns radiation. One way to approach this problem is to first add dissipation to FFE and calculate the dissipative pulsar magnetosphere (and then find the microscopic mechanism of the finite conductivity and the radiation counterpart of the dissipation). 

Strong-Field Electrodynamics (SFE), a dissipative extension of FFE, has already been formulated (Gruzinov 2007 b). It was shown that in the high conductivity limit, in the regions where the ideal FFE 4-current is space-like, SFE reduces to FFE with small dissipative corrections. 

On the other hand, in the regions where the current is time-like we found finite-time singularities. Because of the singularities, we {\it erroneously} concluded that SFE reduces to FFE only in the space-like current regions, and fails in the time-like current regions. 

The presence of singularities and the failure of the model are not equivalent \footnote{Here I have to quote Jeremy Goodman, ``...I'm not sure that this constitutes a ``failure'' of SFE....there are well-defined weak solutions, i.e. solutions with discontinuities ...''.}. The singularities can be physical, like shocks and tangential discontinuities in hydrodynamics and current layers in FFE.  Indeed we will argue in this paper that SFE (regularized by an arbitrarily small diffusion) is well-defined everywhere. One may directly use SFE to study dissipation in FFE. In particular, it should be possible to use SFE to calculate the dissipative pulsar magnetosphere.

It remains to be seen if SFE will be useful in calculating the pulsar radiation, but one thing is already clear. If SFE gives a correct description of dissipative corrections to FFE, the emission of bright pulsars must primarily come from the magnetic separatrix, where the 4-current is space-like (and infinite in the FFE approximation).

\section {FFE and SFE equations}

FFE describes electromagnetic fields of special geometry, with electric field smaller than and perpendicular to the magnetic field. In 3+1 formulation, FFE is defined by Maxwell equations with the ideal Ohm's law 
\begin{equation}\label{FFE3}
\partial _t{\bf B} =-\nabla \times {\bf E},~~~ \partial _t{\bf E}=\nabla \times {\bf B}-{\bf j},~~~{\bf j}={({\bf B}\cdot \nabla \times {\bf B}-{\bf E}\cdot \nabla \times {\bf E}){\bf B}+(\nabla \cdot {\bf E}){\bf E}\times {\bf B} \over B^2}.
\end{equation}
Covariant form of FFE is 
\begin{equation}\label{FFE4}
\partial _\nu F^{\mu \nu}=-j^\mu , ~~~ F^{\mu \nu}j_\nu =0. 
\end{equation}
$F$ is the electromagnetic field tensor. 

FFE equations come from the following physics. At each event go to a {\it good} frame, where ${\bf E}$ is parallel to ${\bf B}$. In FFE,  in a {\it good} frame, the electric field actually vanishes and the current flows along the magnetic field. This physical formulation of FFE has been used to argue that FFE works for Crab-like pulsars. 

SFE describes electromagnetic fields of arbitrary geometry. In 3+1 formulation, SFE is defined by Maxwell equations with the Ohm's law
\begin{equation}\label{ohmsfe}
{\bf j}={\rho {\bf E}\times {\bf B} +(\rho ^2+\gamma ^2\sigma ^2E_0^2)^{1/2}(B_0{\bf B}+E_0{\bf E})\over B^2+E_0^2},
\end{equation}
where 
\begin{equation}\label{gamma}
B_0^2-E_0^2\equiv {\bf B}^2-{\bf E}^2, ~~~ B_0E_0\equiv {\bf E}\cdot {\bf B}, ~~~E_0\geq 0,~~\gamma ^2 \equiv {B^2+E_0^2\over B_0^2+E_0^2},
\end{equation}
$\rho \equiv \nabla \cdot {\bf E}$ is the charge density, and the conductivity scalar $\sigma=\sigma(B_0,E_0)$ is an arbitrary function of the field invariants. The covariant formulation of SFE is
\begin{equation}\label{SFE4}
\partial _\nu F^{\mu \nu}=-j^\mu , ~~~ B_0F^{\mu \nu}j_\nu =E_0\tilde{F}^{\mu \nu}j_\nu , ~~~ j_\mu j^\mu =-\sigma^2E_0^2.
\end{equation}
$\tilde{F}$ is the dual tensor. 

SFE equations come from the following physics. At each event, the {\it best} of the {\it good} frames exists, where ${\bf E}$ is parallel to ${\bf B}$ and the charge density vanishes. In the {\it best} frame, the current $\sigma E$ flows along the magnetic field.

\section{SFE in action}

\subsection{Is SFE well-defined?}

The SFE Ohm's law is dissipative, ${\bf E}\cdot {\bf j}\geq 0$ -- the electromagnetic energy damping is positive or zero in any frame. One can also show that ultraviolet modes propagate on or within the light cone. These good properties should be taken as an indication that SFE is well-defined.

However, in (Gruzinov 2007b) we found finite-time singularities for certain initial fields (infinite second derivative of the electric field, to be discussed in \S 3.3). We erroneously concluded that SFE fails for such initial fields. 

Here we show that these singularities are physical, like shocks in compressible hydrodynamics. It appears that SFE is actually well-defined. We have confirmed this by a large number of numerical simulations of SFE, which showed no pathologies, in the sense that the evolution of the field was independent of the regularization.

\subsection{Regularized SFE}
The most straightforward regularization was used in Maxwell equations (realized as the Lax scheme in numerical simulations):
\begin{equation}\label{regmax}
\partial _t{\bf B} =-\nabla \times {\bf E}+0\cdot \Delta {\bf B},~~~ \partial _t{\bf E}=\nabla \times {\bf B}-{\bf j}+0\cdot \Delta {\bf E}. 
\end{equation}
The SFE Ohm's law needs regularization only in the regions where both invariants of the electromagnetic field vanish. We have used the following regularization when calculating $B_0$ and $E_0$:
\begin{equation}\label{regohm}
B_0^2={B^2-E^2+\sqrt{ (B^2-E^2)^2+4({\bf E}\cdot {\bf B})^2}\over 2}+0,
\end{equation}
and then calculated $E_0$ and $B_0$ from unmodified expressions
\begin{equation}
E_0=\sqrt{ B_0^2-B^2+E^2},~~~B_0=sign({\bf E}\cdot {\bf B})\sqrt{B_0^2}.
\end{equation}
The regularization of the $sign$ function is arbitrary.

\begin{figure}
\epsscale{1.}
\plotone{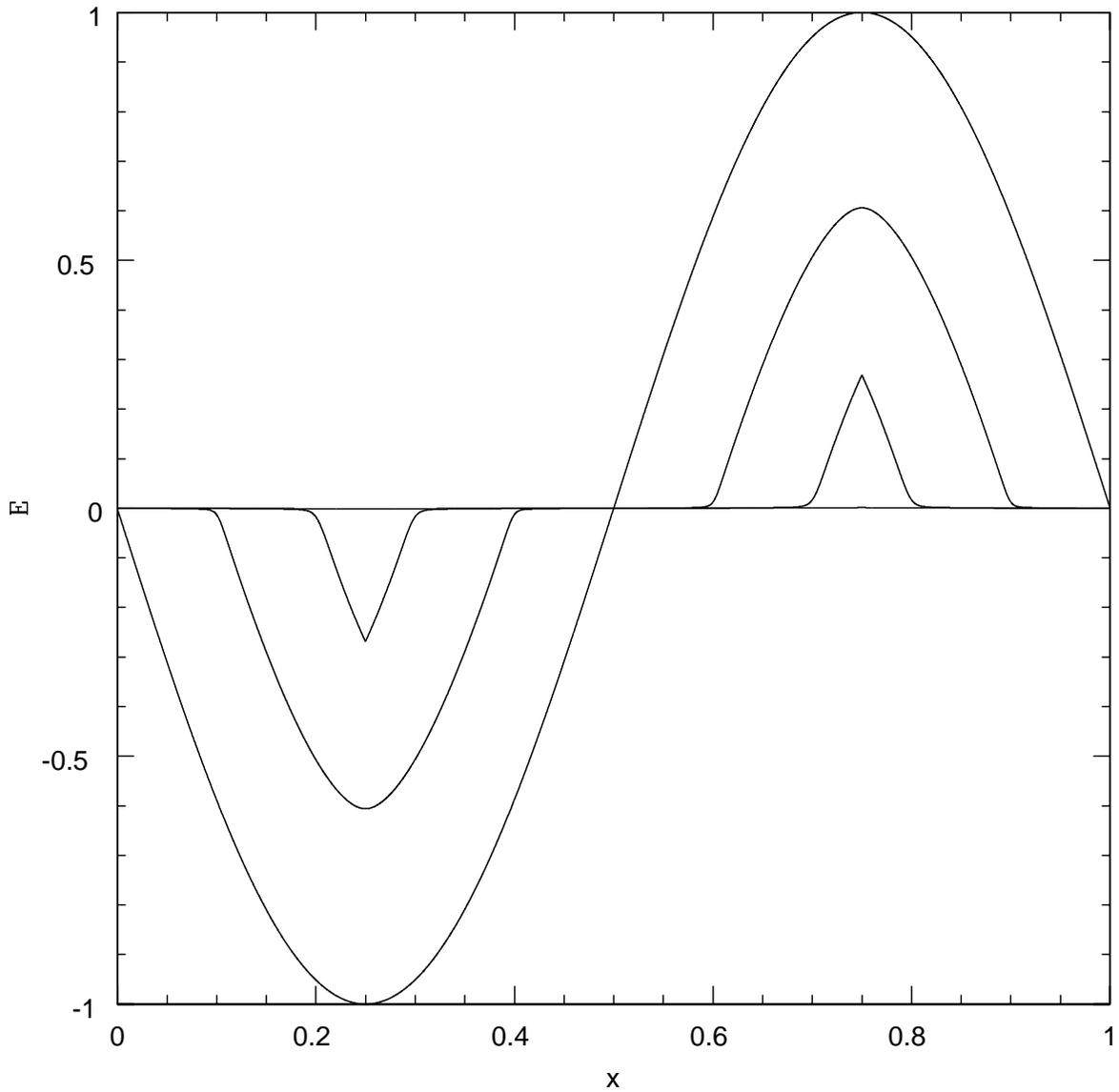}
\caption{Charge relaxation in 1D SFE. The electric field is shown at the initial moment (sinusoidal) and at three later moments (lower amplitude curves). Singularity immediately forms at zero electric field. Singularities also form at the extrema of the electric field.}
\end{figure}

\subsection{Charge relaxation in 1D}
To show both regularizations in action, consider one-dimensional charge relaxation. That is take an initial field ${\bf B}=0$, ${\bf E}=(E(x),0,0)$. Then only the x-component of the electric field $E$ will be non-zero, and SFE equations reduce to
\begin{equation}\label{rel}
\dot{E}=-sign(E)\sqrt{E'^2+\sigma^2 E^2},
\end{equation}
where dot and prime are the time and $x$ derivatives. Here and below, we take $\sigma =const$.

It is easy to show that equation (\ref{rel}) gives finite-time singularities near the extrema of $E$. Say there is a positive maximum of the initial field at $x=0$. Then at all time $t>0$, the Taylor expansion near $x=0$ is $E=a(t)-b(t)x^2+o(x^2)$ with positive $a$, $b$, and equation (\ref{rel}) gives
\begin{equation}
\dot{a}=-\sigma a, ~~~ \dot{b}=2{b^2\over \sigma a}-\sigma b.
\end{equation}
It follows that $b$ becomes infinite in a finite time. At this moment the second derivative of $E$ becomes infinite. 

But so what? The regularized (according to \S 3.2) SFE equations keep working as shown in fig. 1. In fact, we also see that singularities immediately form near the electric field zeros. The electric field zeros develop into zero-field regions bounded by electric field singularities (again infinite second derivative of the field). The electric field dies out in a finite time, even if the conductivity is zero. In SFE, the damping rate  
\begin{equation}
q\equiv {\bf j}\cdot {\bf E}=E_0\sqrt{\rho ^2+\gamma ^2\sigma ^2E_0^2}
\end{equation}
can be positive even for zero conductivity $\sigma$.

\subsection{1D SFE. The FFE limit}
Generic 1D SFE describes electromagnetic fields ${\bf B}=(0,B_y,B_z)$, ${\bf E}=(E_x,E_y,E_z)$, where all components are functions of $x$ and $t$. We ran many simulations of 1D SFE. It appears that SFE (regularized according to \S 3.2) works in 1D.

It is interesting to see what happens to 1D FFE equilibria in the SFE description. The field 
\begin{equation}\label{ffeeq}
E_y=0, ~~ E_z=0, ~~B_y^2+B_z^2-E_x^2=const>0
\end{equation}
is an FFE equilibrium. We used the field (\ref{ffeeq}) as an initial condition for numerical SFE. It was found that equilibria with the space-like 4-current 
\begin{equation}
j^2=E_x'^2-B_y'^2-B_z'^2<0
\end{equation}
decay, and the decay rate decreases with the increasing conductivity $\sigma$. It was also found that equilibria with the time-like 4-current, $j^2>0$, are not damped at all. 

The origin of this {\it semi-dissipative} behavior is clear. In the space-like case, a non-zero parallel electric field is needed to sustain the current. The dissipation rate $q$ is then positive. The dissipation rate decreases for increasing $\sigma$, because for a given space-like current, $E_0=\sqrt{-j^2}/\sigma$ decreases. 

In the time-like current regions, the parallel electric field oscillates around zero, giving a time-like time-averaged current. The amplitude of these oscillations is infinitesimal, and the dissipation $q$ vanishes. 

\subsection{3D SFE}
We ran many simulations of full 3D SFE in a periodic cube, with resolution up to $100^3$. It appears that SFE (regularized according to \S 3.2) works in 3D too. 

We checked, for example, that a space-like FFE equilibrium $E=0$, ${\bf B}=(\sin y-\cos z,~\sin z-\cos x,~\sin x-\cos y)$ is damped and the damping rate goes down for increasing $\sigma$. We also observed formation of zero-field regions and infinite second derivatives for a spherically symmetrical pure electric field.

\acknowledgements
I thank Jeremy Goodman who pointed out that SFE works better than I originally thought. 

This work was supported by the David and Lucile Packard foundation.

\end{document}